\documentclass[conference]{IEEEtran}
\IEEEoverridecommandlockouts
\usepackage{amsmath,amssymb,amsfonts}
\usepackage{graphicx}
\usepackage{textcomp}
\usepackage{xcolor}
\usepackage{comment}
\usepackage{algorithm}
\usepackage[noend]{algpseudocode}
\usepackage{xcolor}
\usepackage{fontawesome}
\usepackage{tikz}
\usepackage{comment}
\usepackage{centernot}

\usepackage{subfig,hyperref}
\def\BibTeX{{\rm B\kern-.05em{\sc i\kern-.025em b}\kern-.08em
    T\kern-.1667em\lower.7ex\hbox{E}\kern-.125emX}}
\begin{document}

\title{Automated Consistency Analysis of LLMs\\
}

\author{\IEEEauthorblockN{Aditya Patwardhan}
\IEEEauthorblockA{\textit{Department of Computer Science} \\
\textit{Stony Brook University}\\
Stony Brook, USA \\
aapatwardhan@cs.stonybrook.edu}
\and
\IEEEauthorblockN{Vivek Vaidya}
\IEEEauthorblockA{\textit{Department of Computer Science} \\
\textit{Rutgers University}\\
New Brunswick, USA \\
vivek.vaidya@rutgers.edu}
\and
\IEEEauthorblockN{Ashish Kundu}
\IEEEauthorblockA{\textit{Cisco Research} \\
San Jose, USA \\
ashkundu@cisco.com}
}

\maketitle

\begin{abstract}
Generative AI (Gen AI) with large language models (LLMs) are being widely adopted across the industry, academia and government. Cybersecurity is one of the key sectors where LLMs can be and/or are already being used. There are a number of problems that inhibit the adoption of trustworthy Gen AI and LLMs in cybersecurity and such other critical areas. One of the key challenge to the trustworthiness and reliability of LLMs is: how consistent an LLM is in its responses? In this paper, we have analyzed and developed a formal definition of consistency of responses of LLMs. We have formally defined what is consistency of responses and then develop a framework for consistency evaluation. The paper proposes two approaches to validate consistency: self-validation, and validation across multiple LLMs. We have carried out extensive experiments for several LLMs such as GPT4oMini, GPT3.5, Gemini, Cohere, and Llama3, on a security benchmark consisting of several cybersecurity questions: informational and situational. Our experiments corroborate the fact that even though these LLMs are being considered and/or already being used for several cybersecurity tasks today, they are often inconsistent in their responses, and thus are untrustworthy and unreliable for cybersecurity.

\end{abstract}


\section{Introduction}
Over the last couple of years, the rapid development of Generative AI, and wide adoption of Large Language Models (LLMs) have revolutionized the field of natural language processing, and automated  generation of content in several languages, in  multi modal manner such as text, audio, and video. These machine learning and AI models, related AI agents~\cite{talebirad2023multi} are also being widely used for automation of tasks across the industry and several sectors. One such important sector is cybersecurity. LLMs have already been used to assist with cybersecurity processes (SecOps)~\cite{gennari2024considerations}, in security operation centers (SoC)~\cite{saha2024llm}, in security analysis of code and configurations~\cite{toth2024llms, minna2024analyzing}, in generation of secure code~\cite{vaidya2023critical,saha2024empowering}, in security and penetration testing~\cite{song2024poster} among many others. 

Trustworthiness of such LLMs and generative AI is a critical factor in whether and how we use it in cybersecurity as well as other real-world applications and systems. LLMs suffer from security issues~\cite{wu2024new,qachfar2024all}, and there are risks of code generation using LLMs~\cite{vaidya2023critical}. LLMs also hallucinate~\cite{xu2024hallucination} and such hallucinations often increase the risk of reduction of utility of the models, or even risk of failure during the task execution with LLMs as judges~\cite{zheng2024judging}, or autonomous LLM agents~\cite{talebirad2023multi}. Consistency has a strong relation to hallucination. 

In order to be able to trust the large language models, one key factor is consistency.  Studies of LLM consistency lead to the following questions:  (1) are they reliably consistent, in that can we rely on the responses - are they semantically consistent? And that leads to the following questions further: (2) What is consistency, and (3) how can consistency be evaluated in an automated manner. In this paper, we study these research questions and attempt to address them. Our focus is on semantic consistency of responses by one LLM or a set of LLMs. There are other types of consistency requirements such as: syntactic consistency, structural consistency, which are often in the context of copyrights, and are out of the scope of this paper. 

{\bf Our Contributions};
We have studied the problem of consistency of responses by LLMs especially in the context of cybersecurity. That is because, in such a domain, consistency of responses is quite important for making correct and effective decisions, otherwise inconsistency may lead to security vulnerabilities, weaknesses and/or liabilities and harm to systems, users and to enterprises. We have formally defined what is consistency of responses and then develop a framework for consistency evaluation. The paper proposes two approaches to validate consistency: self-validation, and validation across multiple LLMs. We have carried out extensive experiments for several LLMs such as ChatGPT 4o Mini, GPT3.5, Gemini, Cohere, and Llama3, on a security benchmark consisting of both informational and situational cybersecurity questions. Our experiments corroborate the fact that even though these LLMs are being considered and/or already being used for several cybersecurity tasks today, they are often inconsistent in their responses. Thus unless the consistency of LLMs in their responses is improved to a reliable level, they cannot be trusted to an extent that they can be used in enterprise-level cybersecurity operations. 

{\bf Organization of the paper}: Section~\ref{sec:def} defines consistency of LLM responses formally and analyzes its relationship with accuracy and hallucination.  Section~\ref{sec:framework} describes the framework of self-validation and cross-validation of consistency and the algorithms. In Section~\ref{sec:experiments} we have presented the experimental benchmark, settings and analyzed the experimental results. Section~\ref{sec:related} discusses the related work, and Section~\ref{sec:conclusions} concludes the paper with future work.

\section{Consistency in the context of LLMs} \label{sec:def}
One key question arises on the use of LLMs is - can we trust an LLM's responses? Reliability of an LLM is dependent on its consistency. 

\subsection{Consistency of Responses}
Consistency of LLMs is defined based on how consistent an LLM's responses to different prompts are. It is about whether the responses that an LLM returns to the same or semantically identical prompts are sufficiently similar or identical to each other. Such responses may be syntactically or structurally different but semantically identical. If a set of responses are not sufficiently similar or identical to each other then we refer to it as "inconsistent responses".   

The prompts are processed by the LLM over a duration of time $\Delta_t$, during which the LLM model is assumed to remain stable and unchanged. In this paper, by the clause - "responses of an LLM to a prompt" means "responses of an LLM to the same prompt or semantically identical prompts issued multiple times". Each issuance of a prompt to an LLM is called as a query.

\subsection{Consistency and other Properties}
Some of the key implications of consistency and inconsistency in the context of LLMs are as follows:
\begin{itemize}    
    \item Accuracy $A$ $\implies$ Consistency $C$: if an LLM is stated to  be accurate in its responses to a prompt, then such responses will be consistent with respect to each other. In turn, lack of consistency implies lack of accuracy: $\neg C \implies \neg A$. 
    \item Consistency $C$ $\centernot\implies$ Accuracy $A$: Semantic consistency of responses does not always imply accuracy of the responses, primarily because an LLM may respond with the semantically equivalent responses for the same prompt or semantically equivalent prompts each time it is queried, but guaranteeing the semantic accuracy of the responses is beyond the problem of maintaining consistency.
    \item Consistency of LLMs are related to the hallucination of LLMs~\cite{mcdonald2024reducing}: The more inconsistent an LLM is in its responses, the more the LLM may hallucinate simply because the LLM is responding with semantically different responses to the same prompt over multiple queries. Section~\ref{sec:hallucination} presents some related findings on hallucination. However, a detailed discussion and analysis of such a relationship is beyond the scope of this paper.
\end{itemize}

\subsection{Formal Definition}

Let $L_i$ refer to a large language model (LLM). Let $p_k$ refer to a prompt in one or more languages that is/are supported by an $L_i$. Let $s_j$ represent an active user session of a $w$'th user entity $u_w$. Let $t_l$ represent the point in time at which a query is made to an LLM or the time at which an LLM responds to a query such that the response is complete (not in the process of incremental output). 
Let $r_v$ $\xleftarrow{}$ $q_v(L_i, s_j, p_k, t_l, u_w)$ refer to a unique query the user $u$ using prompt $p_j$ to $L_i$ leading to response $r_v$. The terms "Prompt" and "Query" are used interchangeably throughout this paper.

Let $L_i$ $\in$ $\mathcal{L}$, which is a set of  LLMs, where $L_i$ $\in$ $\mathcal{L}$, refer to a large language model (LLM). Let $R_v$ $\xleftarrow{}$ $Q_v(\mathcal{L}_i, S_j, P_k, t_l, U_w)$ refer to a unique set of queries the user $u$ using prompt $p_j$ to a subset of LLMs $\mathcal{L}_i$ $\subseteq$ $\mathcal{L}$ leading to a set of responses $R_v$ that are received by the set of users $U_w$, over a set of active sessions $S_j$, where there is an one-to-one mapping of all the elements across the sets of responses, queries, LLMs, sessions, prompts, timestamps of completion of queries, and users.

Prompts $p_x$ and $p_y$ are semantically equivalent, i.e. they are identical, sufficiently identical or sufficiently similar to each other semantically, which is represented by $p_x$ $\equiv$ $p_y$. Responses $r_v$ and $r_w$ are semantically equivalent, i.e. they are identical, sufficiently identical, or sufficiently similar to each other semantically, which is represented by $r_v$ $\equiv$ $r_w$.\\ 

\textbf{Definition 1: Consistency of Responses -- One LLM.}\\
    Given two queries $r_1$  $\xleftarrow{}$ $q_1(L_1, s_1, p_x, t_1, u_1)$ and $r_2$  $\xleftarrow{}$ $q_2(L_1, s_1, p_y, t_3, u_1)$, where $p_x \equiv p_y$, the responses are consistent if $r_1$ $\equiv$ $r_2$.\\ 

\textbf{Definition 2: Consistency of Responses -- Multiple LLMs.}\\ 
    Given two queries $R_1$  $\xleftarrow{}$ $Q_1(\mathcal{L}_1, S_1, p_x, T_1, u_1)$ and $R_2$  $\xleftarrow{}$ $Q_2(\mathcal{L}_1, S_1, p_y, T_3, u_1)$, where $p_x \equiv p_y$, the responses are consistent if for all pairs $<d_v, e_v>$, where $d_v \in R_1$  and $e_v \in R_2$,  $x_v$ $\equiv$ $y_v$.

In the rest of the paper, on the basis of this formal model, we will study and carry out experiments on black-box LLMs.

\section{Consistency Validation Framework} \label{sec:framework}

In this paper, we propose a comprehensive evaluation framework that incorporates multiple algorithms for evaluating consistency and accuracy, providing a holistic metric of how trustworthy an LLM is. In this paper, we measure LLM consistency in the context of cybersecurity applications.

The rest of the paper follows the formal model and definitions. However, the notations used may differ in order to provide more readability for the algorithms and discussion.

\subsection{Consistency}

To be trustworthy, an LLM has to return a similar answer every time it’s prompted with the same question, so different users don’t get different answers or explanations to answers when researching the same topic. Our consistency algorithm gives an LLM the same prompt $n$ times and evaluates the similarity between responses using multiple metrics such as Jaccard Index~\cite{article}, Cosine Similarity~\cite{cosine_similarity}, Sequence Matcher~\cite{python_doc}, and Levenshtein distance~\cite{levenshtein_distance}, all standardized to a scale of 0 to 100. 

The Consistency algorithm (Algorithm \ref{alg:consistency}) operates in three modes: low, medium, and high, where higher settings require progressively greater consistency in the metrics for the model to be considered consistent. For each question, the algorithm collects $k$ model responses and then calculates pairwise consistency scores using the four metrics for every possible pair of responses, including consecutive responses. If the metric score is higher than a certain threshold, that pair passes for that metric. Therefore, while consecutive comparisons are part of the pairwise evaluation, the algorithm ensures a comprehensive assessment by comparing all responses in the set. 

If the pair passes $x$ out of 4 consistency score metrics, it is considered to pass overall. If 80\% of pairs pass, the model is considered consistent for that question. If 80\% of questions pass, the model is considered consistent overall.

Instead of keeping the same percentages to pass each metric, we have implemented low, medium, and high settings to further bring out the differences between the models. Under these settings, the percentage required for a pair to "pass" a certain consistency metric changes. 

For the low threshold, Jaccard and Cosine have to be 70\%, and Sequence Matcher and Levenshtein have to be 20\%. For medium, Jaccard and Cosine have to be 80\%, and Sequence Matcher and Levenshtein have to be 40\%. For high, Jaccard and Cosine have to be 90\%, and Sequence Matcher and Levenshtein have to be 60\%. Sequence Matcher and Levenshtein Distance similarity take the order of characters into account as opposed to the two, so they tend to be more critical of responses that are roughly the same but worded differently. Due to this, their required percentages are significantly lower than the other two.

\begin{algorithm}[tb]
\caption{Consistency Analysis}
\label{alg:consistency}
\begin{algorithmic}[1]
\Statex \textbf{Input:} \textit{LLM} $L_i$ -  LLM to perform consistency analysis
\Statex \textbf{Input:} \textit{Prompts/Queries} -  list of queries to be validated
\Statex \textbf{Input:} \textit{k} - The number of repetitions for validation
\Statex \textbf{Input:} \textit{simthreshold} - The threshold checked to determine similarity: low, medium, high
\Statex \textbf{Input:} \textit{qthreshold} - The minimum fraction of questions for which the LLM's answers need to be consistent
\Statex
\Statex \textbf{Output:} True/False
\Statex
\Procedure{Consistency\_Analysis}{\textit{LLM, Queries, k, simthresh, qthreshold}}
\State $qcnt \gets 0$
\State $npt \gets 0.8 * k * (k-1) / 2$
\For{each $q \in Queries$}
  \State $Resp \gets [~]$
  \For{$i \in 1 \dots k$}
    \State $Resp_i \gets LLM\_Api(q)$
  \EndFor
  \State $SS\_cnt,LS\_cnt,JS\_cnt,CS\_cnt \gets 0$
  \For{$i \in 1 \dots k-1$}
    \For{$j \in i+1 \dots k$}
        \State $SS \gets SeqMatcher(Resp_i,Resp_j)$
        \State $LS \gets LevenDist(Resp_i,Resp_j)$
        \State $JS \gets JaccardCoef(Resp_i,Resp_j)$
        \State $CS \gets CosineSim(Resp_i,Resp_j)$
        \State $SS\_cnt$ += $SS \ge simthresh$ ? 1 : 0
        \State $LS\_cnt$ += $LS \ge simthresh$ ? 1 : 0
        \State $JS\_cnt$ += $JS \ge simthresh$ ? 1 : 0
        \State $CS\_cnt$ += $CS \ge simthresh$ ? 1 : 0
    \EndFor
  \EndFor
  \If{$SS\_cnt,LS\_cnt,JS\_cnt,CS\_cnt \ge npt$}
    \State $qcnt \gets qcnt + 1$
  \EndIf
\EndFor
\If{$qcnt / |Queries| \ge qthreshold$}
  \State \Return{$true$}
\Else
  \State \Return{$false$}
\EndIf
\EndProcedure
\end{algorithmic}
\end{algorithm}

\subsection{Agreement}
To be trustworthy an LLM has to return the correct answer to a question. To determine if the LLMs agree on whether a certain answer is correct or not, our framework uses two algorithms. The first is Self-Validation, where an LLM checks it's own answer to a question. The second is Cross-Validation, where an LLM's answer to a question is checked by every other LLM. An LLM must be considered accurate by both these algorithms to be considered trustworthy in terms of information accuracy.

\subsubsection{Self-Validation}

\begin{figure}[t]
    \centering
    \includegraphics[width=\linewidth]{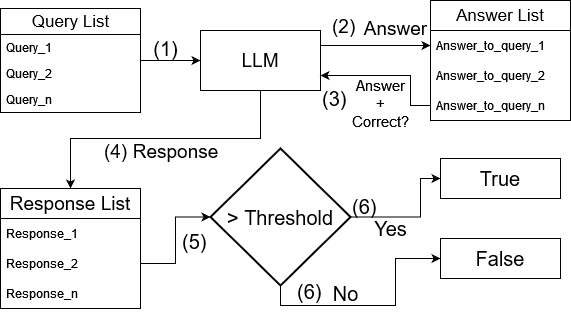}
    \caption{Self-Validation Architecure}
    \label{fig:Self-Valid-Alg-Fig}
\end{figure}

Figure~\ref{fig:Self-Valid-Alg-Fig} illustrates the Self-Validation framework for Large Language Models (LLMs). In this process, the LLM generates a "Response List" with repeated responses to the same question. That same LLM is then asked whether the generated responses are the correct answer to the original query. If it agrees with enough of its own responses, it is considered factually consistent by self-validation.

The Self-Validation Algorithm (Algorithm \ref{Self Validation}) has an LLM to evaluate the accuracy of its answers as shown in \ref{fig:Self-Valid-Alg-Fig} . It prompts the LLM with its answer to a question and asks if it is the correct answer to that question. This is done $k$ times for every question. If the LLM responds "yes" 80\% of the time, that question is considered correct by this metric. If the number of correct questions divided by the total number of questions is greater than $qthreshold$, the LLM is considered accurate overall by this metric.

\begin{algorithm}[tb]
\caption{Self Validation}
\label{Self Validation}
\begin{algorithmic}[1]
\Statex \textbf{Input:} \textit{LLM} $L_i$ - The LLM to validate
\Statex \textbf{Input:} \textit{Queries} - The list of queries to be validated
\Statex \textbf{Input:} \textit{k} - The number of repetitions for validation
\Statex \textbf{Input:} \textit{qthreshold} - The minimum fraction of questions for which the LLM needs to accept its own answer to be considered successful
\Statex
\Statex \textbf{Output:} True/False
\Statex
\Procedure{Self\_Validation}{\textit{LLM, Queries, k, qthreshold}}
\State $qcnt \gets 0$
\For{each $q \in Queries$}
  \State $Orig\_Resp \gets LLM\_Query\_Api(q)$
  \State $svq \gets $q + Orig\_Resp + ``correct? yes or no''
  \State $valcnt \gets 0$
  \For{$i \in 1 \dots k$}
    \State $Resp_i \gets LLM\_Api(svp)$
    \If{$Resp_i$ = Yes}
        \State $valcnt \gets valcnt + 1$
    \EndIf
  \EndFor
  \If{$valcnt > 0.8 * k$}
    \State $qcnt \gets qcnt + 1$
  \EndIf
\EndFor
\If{$qcnt / |Queries| \ge qthreshold$}
  \State \Return{$true$}
\Else
  \State \Return{$false$}
\EndIf
\EndProcedure
\end{algorithmic}
\end{algorithm}

\subsubsection{Cross-Validation}

\begin{figure}[!t]
    \centering
    \includegraphics[width=\linewidth]{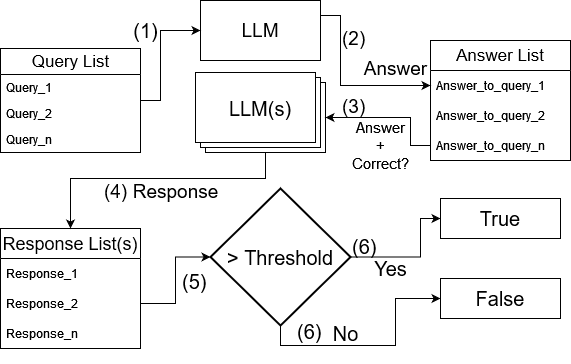}
    \caption{Cross-Validation Architecture}
    \label{fig:Cross-Valid-Alg-Fig}
\end{figure}

\ref{fig:Cross-Valid-Alg-Fig} illustrates the cross-validation framework for evaluating the consistency of a Large Language Model (LLM). This framework fact checks LLM responses with other LLMs. Each LLM generates a response to a prompt, and then every other LLM is asked whether that LLM's response to the original prompt is correct. If there is enough agreement between LLMs, that LLM is considered factually consistent by cross-validation

If an LLM has unreliable information caused by biased training data, it may not be able to recognize that in the self-validation step. To remedy that we propose the Cross-Validation Algorithm (Algorithm \ref{fig:Cross-Valid-Alg-Fig}), which cross-validates an LLM's responses with the other LLMs as shown in \ref{fig:cross-valid}. To begin with, the algorithm is provided with a list of one LLMs responses to a set of questions. For each response, all the other LLMs are asked whether it is the correct response to the respective question. If 80\% of the other LLMs' responses are yes to a question, that question is considered correct by this metric. If the number of correct questions divided by the total number of questions is greater than $qthreshold$, the LLM is considered factually consistent overall by this metric.

\begin{algorithm}[tb]
\caption{Cross Validation}
\label{Cross Validation}
\begin{algorithmic}[1]
\Statex \textbf{Input:} \textit{LLMs} $\mathcal{L}$ - The list of LLMs to cross-validate
\Statex \textbf{Input:} \textit{Queries} - The list of queries to be validated
\Statex \textbf{Input:} \textit{k} - The number of repetitions for validation
\Statex \textbf{Input:} \textit{qthreshold} - The minimum fraction of questions for which the other LLMs needs to accept any LLM's answer for it to be considered successful
\Statex
\Statex \textbf{Output:} cv\_llm - a boolean list; $cv\_llm_i$ indicates whether $llm_i$ is cross validated
\Statex
\Procedure{Cross\_Validation}{\textit{LLMs, Queries, k, qthreshold}}
\State $cv\_llm \gets \phi$
\For{each $llm_i \in LLMs$}
    \State $qcnt \gets 0$
    \For{each $q \in Queries$}
      \State $Orig\_Resp \gets llm_i\_Query\_Api(q)$
      \State $svq \gets $q + Orig\_Resp + ``correct? yes or no''
        \State $llmcnt \gets 0$
        \For{each $llm_j \in LLMs$, s.t. $llm_j \neq llm_i$}
          \State $valcnt \gets 0$
          \For{$i \in 1 \dots k$}
            \State $Resp_i \gets LLM_j\_Api(svp)$
            \If{$Resp_i$ = Yes}
                \State $valcnt \gets valcnt + 1$
            \EndIf
          \EndFor
          \If{$valcnt > 0.8 * k$}
              \State $llmcnt \gets llmcnt + 1$
          \EndIf
        \EndFor
        \If{$llmcnt > 0.66 * |LLMs|$}
            \State $qcnt \gets qcnt + 1$
        \EndIf
    \EndFor
    \If{$qcnt / |Queries| \ge qthreshold$}
      \State $cv\_llm_i \gets$ {$true$}
    \Else
      \State $cv\_llm_i \gets$ {$false$}
    \EndIf
\EndFor
\State \Return {cv\_llm}
\EndProcedure
\end{algorithmic}
\end{algorithm}

\section{Empirical Analysis on LLMs using this Framework} \label{sec:experiments}
\subsection{Benchmarks}
To test the LLMs, we developed a benchmark on 40 Cybersecurity interview questions from a popular list of cybersecurity interview questions and answers~\cite{Hiremath_2024}. In the benchmark, the questions are divided into two types as follows.

\subsubsection{Information Questions} 
The first 33 questions are basic information questions, with a well-known correct answer. For example, ``what is cryptography", or ``what is the CIA triad in Cybersecurity"? These should theoretically be the easiest for LLMs to answer and check the answers of. They are listed in Table \ref{tbl:infq}.

\subsubsection{Situation Question}
The last 7 questions are Cybersecurity situation questions. They place the reader in a situation and ask what they should do. For example, "You receive an email from your bank telling you there is a problem with your account. The email provides instructions and a link so you can log into your account and fix the problem. What should you do?" These questions are more open to interpretation and should be harder for LLMs to answer and check. They are listed in Table \ref{tbl:sitq}.

\subsection{Consistency}
Gemini~\cite{geminiteam2024geminifamilyhighlycapable} and Bloom~\cite{workshop2023bloom176bparameteropenaccessmultilingual} are both deterministic models, and return the exact same response when the same prompt is passed. Therefore, they are perfectly consistent for all thresholds. Since their behavior is deterministic, the consistency check does not provide any useful information and they aren't included in the plots. 

\begin{figure}[!t]
    \centering
    \includegraphics[width=\linewidth]{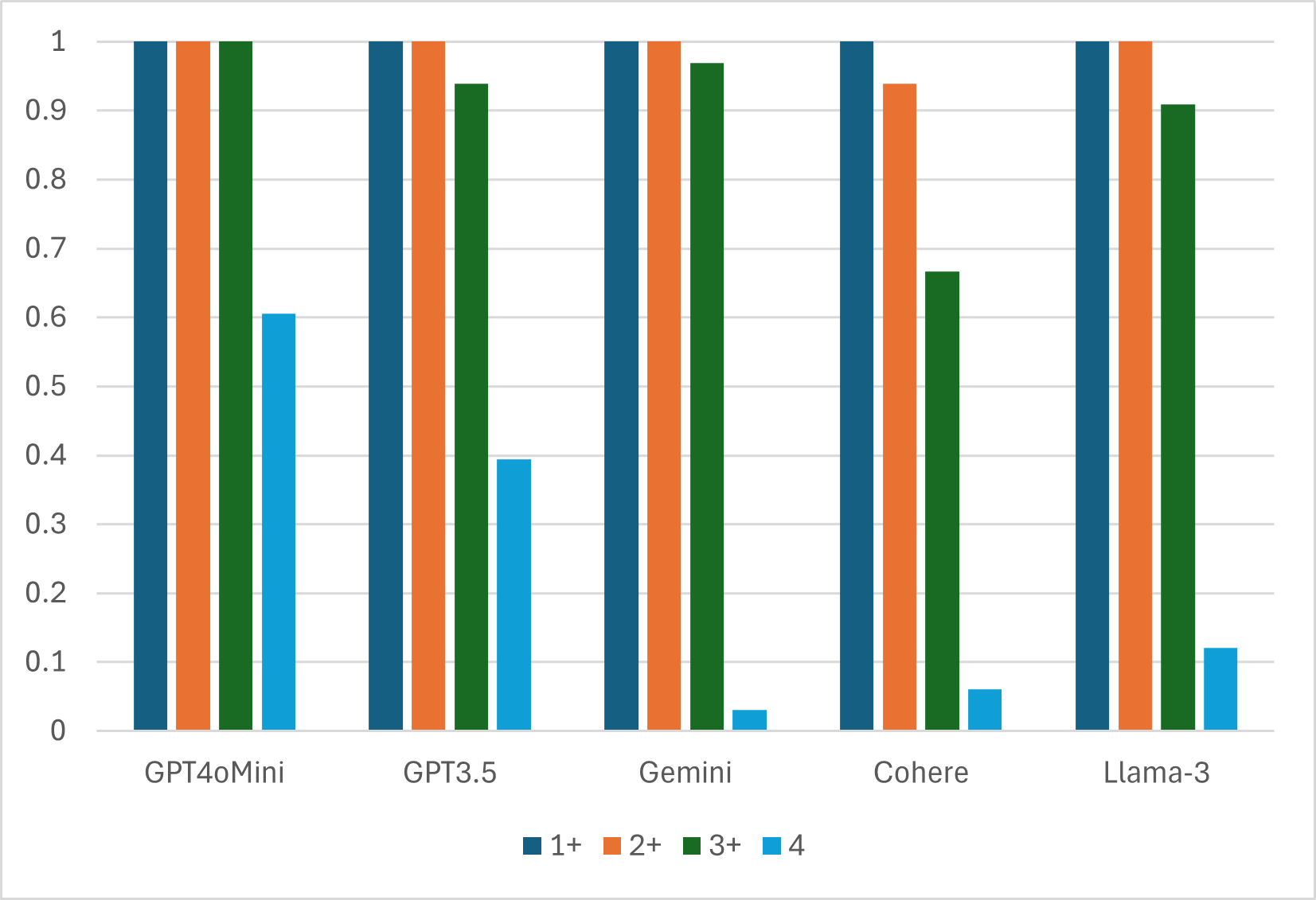}
    \caption{Consistency Analysis for Low threshold}
    \label{fig:consistency-analysis-low}
\end{figure}

\begin{figure}[!t]
    \centering
    \includegraphics[width=\linewidth]{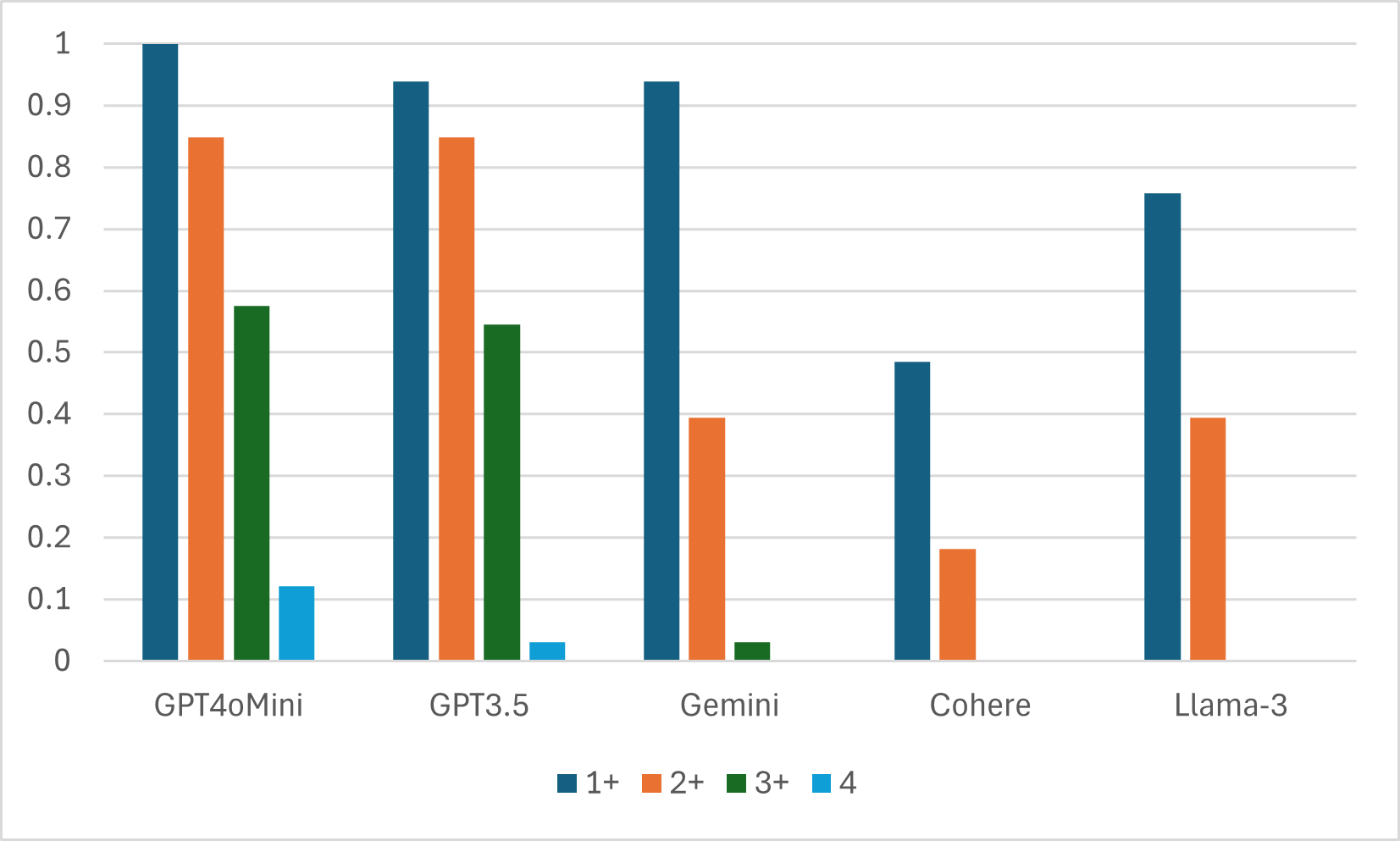}
    \caption{Consistency Analysis for Medium threshold}
    \label{fig:consistency-analysis-med}
\end{figure}

\begin{figure}[!h]
    \centering
    \includegraphics[width=\linewidth]{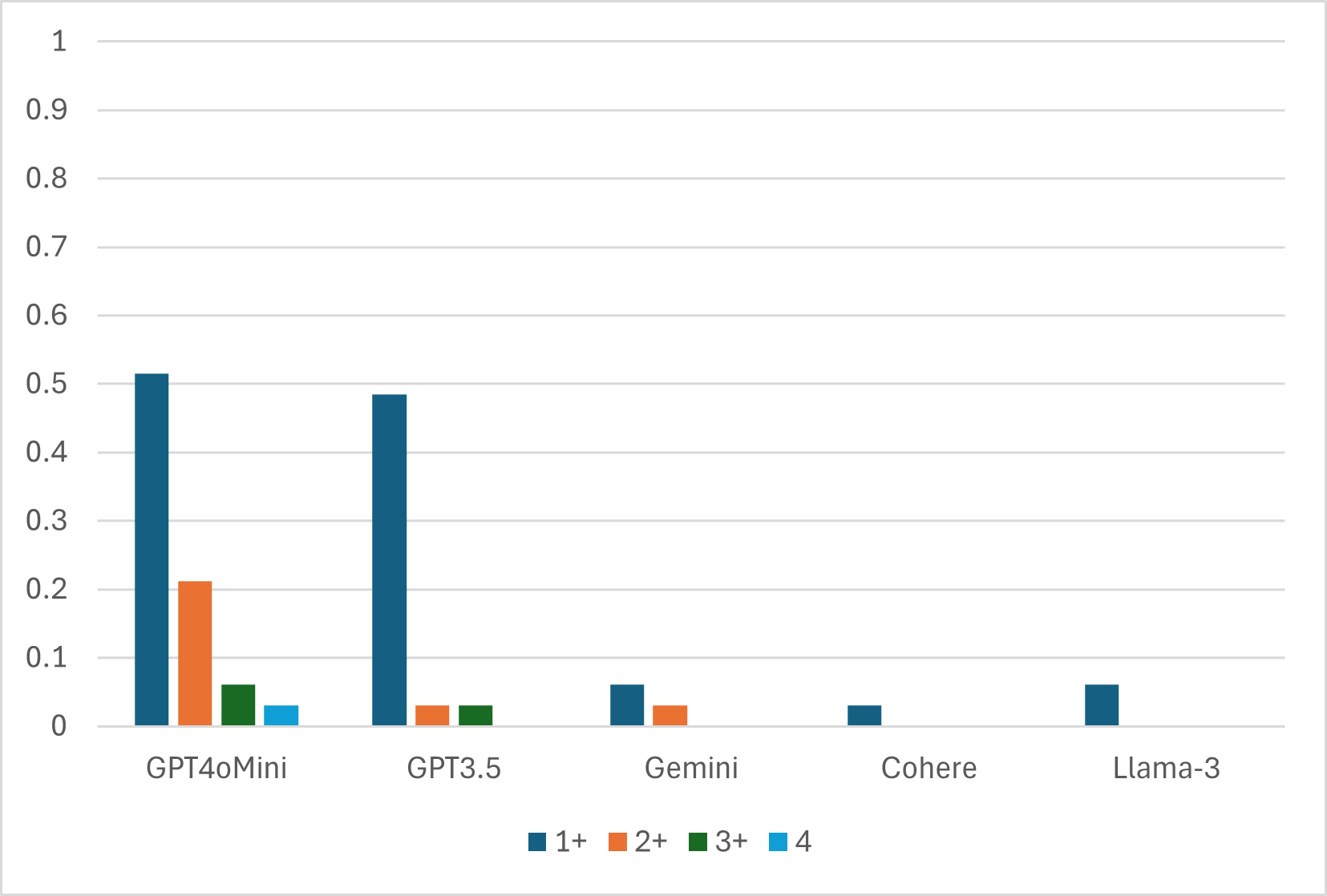}
    \caption{Consistency Analysis for High threshold}
    \label{fig:consistency-analysis-high}
\end{figure}

\begin{figure}[tb]
    \centering
    \includegraphics[width=\linewidth]{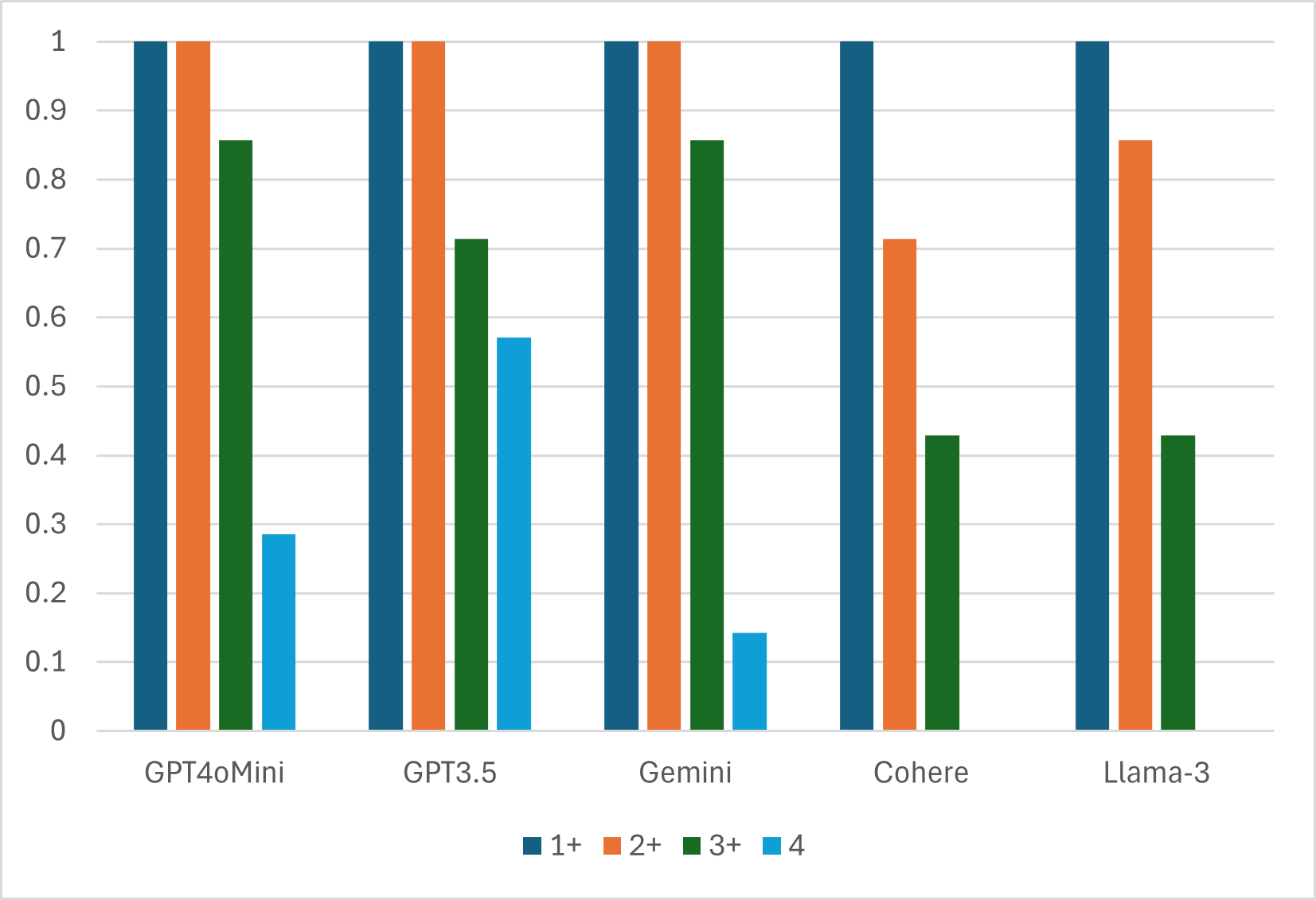}
    \caption{Low threshold for Situational Questions}
    \label{fig:consistency-analysis-low-sit}
\end{figure}

\begin{figure}[tb]
    \centering
    \includegraphics[width=\linewidth]{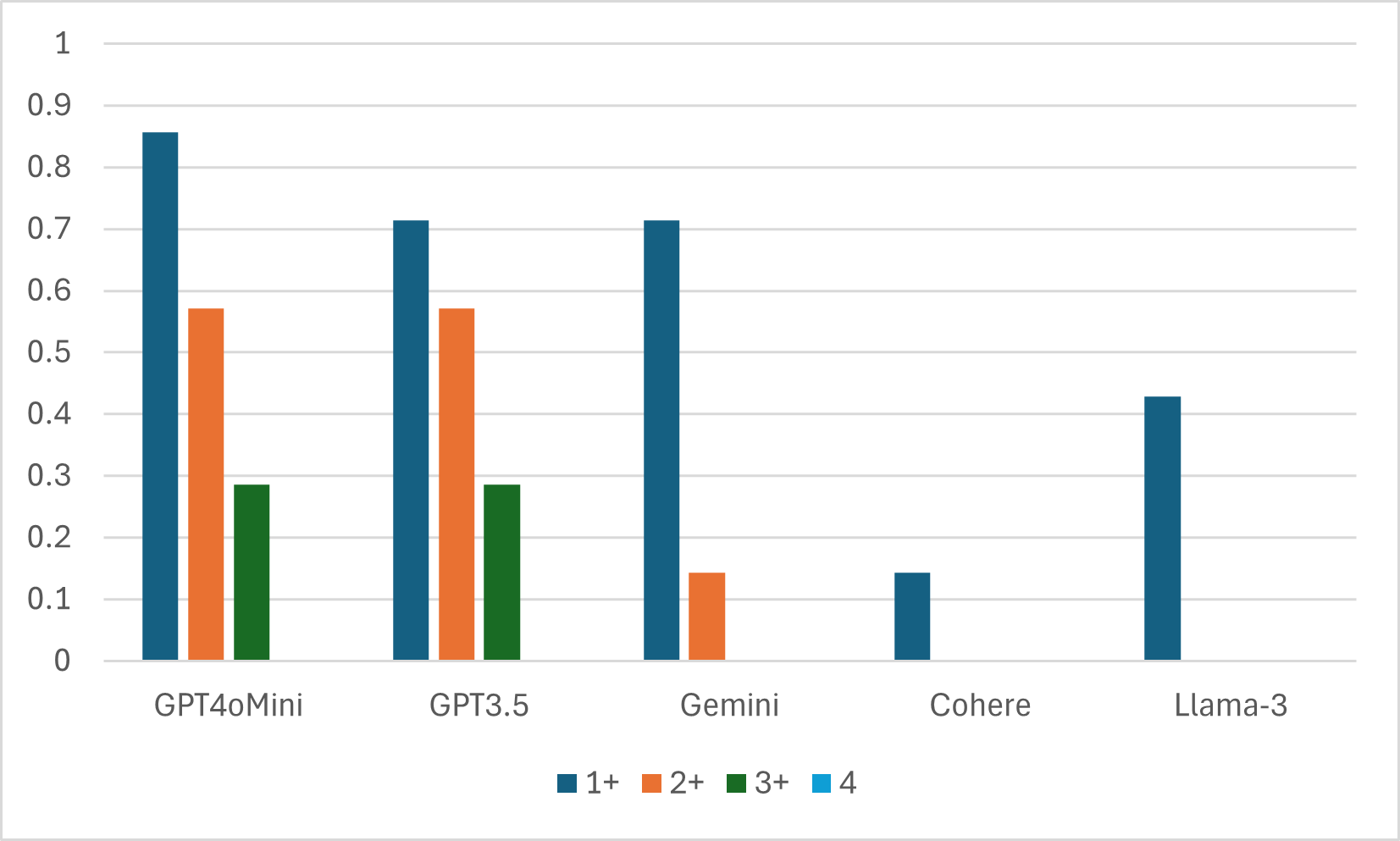}
    \caption{for Medium threshold for Situational Questions}
    \label{fig:consistency-analysis-med-sit}
\end{figure}

\begin{figure}[!h]
    \centering
    \includegraphics[width=\linewidth]{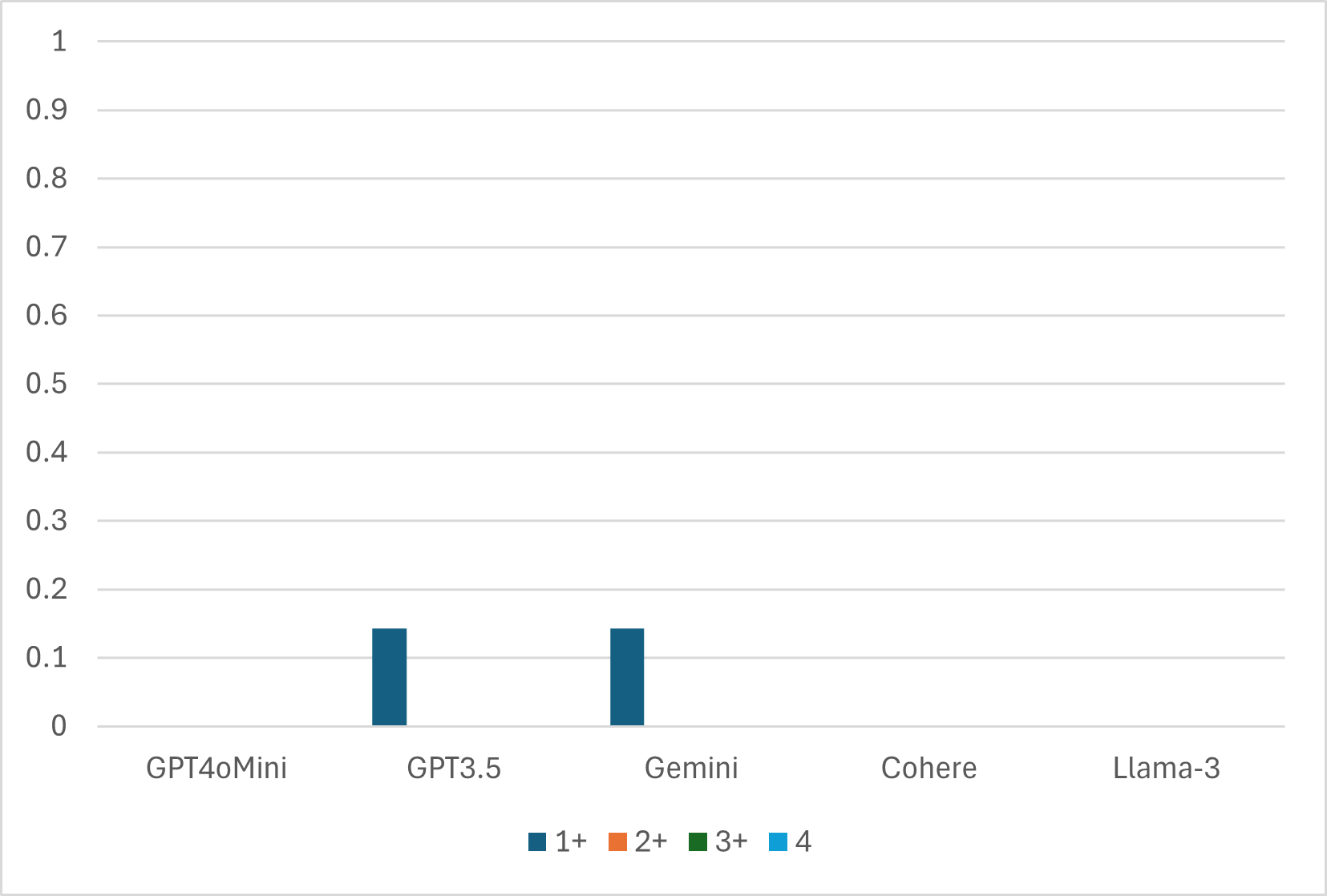}
    \caption{High threshold for Situational Questions}
    \label{fig:consistency-analysis-high-sit}
\end{figure}

Figures \ref{fig:consistency-analysis-low}-\ref{fig:consistency-analysis-high} depict the results for the low, medium, and high thresholds respectively for the 33 information questions. Figures \ref{fig:consistency-analysis-low-sit}-\ref{fig:consistency-analysis-high-sit} provide the corresponding results for the 7 situational questions. In each figure, there are 4 sets of bars corresponding to each LLM, providing the results of the consistency analysis for the cases where 1 or more, 2 or more, 3 or more, or all 4 of the consistency metrics evaluate to true.

Looking at the results for Low threshold with information questions (Figure \ref{fig:consistency-analysis-low}), the majority of LLMs seem to pass when 1-3 similarity metrics are sufficient to pass for a response pair to be considered consistent. When all 4 similarity metrics are needed, the percentage of questions that pass drops drastically, and not a single model is above 80\%. Comparing this with the situational questions (Figure \ref{fig:consistency-analysis-low-sit}), all the results are noticeably lower, as expected. As opposed to all LLMs passing at least once under the Low threshold, none pass under the High threshold. With the information questions (Figure \ref{fig:consistency-analysis-high}), GPT 4o mini performs the best once again, with GPT 3.5 not far behind. 3.5 and 4o are the only ones to have any questions pass when questions have to pass 3/4 metrics, and 4o mini is the only one to have any questions pass when all 4 metrics are used. The difference between the situation questions (Figure \ref{fig:consistency-analysis-high}) and the information questions is also greatest under the High threshold. Interestingly, GPT 3.5 outperforms 4o mini here, and it and Gemini are the only models to have any questions pass when only 1/4 of metrics are used.

The Medium threshold has the most interesting results. With the information questions (Figure \ref{fig:consistency-analysis-med}), GPT 4o Mini and GPT 3.5 are the only models to pass when 2/4 metrics are used. With 1 metric, Gemini passes as well. With the situation questions (Figure \ref{fig:consistency-analysis-med-sit}) GPT 4o Mini was the only model to pass when only 1/4 metrics are used.

Not counting Bloom and Meta OPT~\cite{zhang2022optopenpretrainedtransformer} since they're inherently 100\% consistent, the most consistent models are GPT 4o Mini, GPT 3.5, and Google Gemini, in that order. When putting the average metric scores into tables, in both the regular questions (Table \ref{tab:sim_tab}) and the situation questions (Table \ref{tab:sim_tab}) GPT 4o Mini consistently scores higher in Sequence Matcher and Levenshtein Distance, while 3.5 consistently scores higher on Jaccard Index and Cosine similarity. Sequence and Levenshtein take order into account, so it can be inferred that 4o mini has more variation in the way it words its responses as compared to the older 3.5. Along with that, there is very little variation in the Cosine Similarity and to a lesser degree Jaccard Index scores. This increases a bit on the situation questions but is still considerably lower than the other two. This shows that these two metrics are less useful for comparing LLMs to each other.

\begin{table}[tb]
    \centering
        \caption{Average Similarity Scores per LLM}
    \label{tab:sim_tab}
    \begin{tabular}{|p{1.5cm} | p{1.5cm} | p{1.5cm} | p{1.3cm} | p{1.1cm} |}
        \hline
         Model & Sequence Matcher &  Levenshtein Distance & Jaccard Index & Cosine Similarity \\
         \hline
         GPT 4o Mini	& 30.21 & 46.54 & 89.75 & 89.29 \\ 
         \hline
         GPT 3.5 & 30.82 & 50.56 & 89.49 & 84.63 \\
         \hline
         Gemini & 10.22 & 32.5 & 86.62 & 82.1 \\
         \hline
         Cohere & 13.33 & 33.35 & 79.45 & 81.03 \\
         \hline
         Llama3 & 14.2 & 33.88 & 84.97 & 81.24 \\
         \hline
    \end{tabular}

\end{table}

\begin{table}[tb]
    \centering
        \caption{Average Similarity Scores per LLM Situation Questions}
    \label{tab:sim_tab_sit}
    \begin{tabular}{|p{1.5cm} | p{1.5cm} | p{1.5cm} | p{1.3cm} | p{1.1cm} |}
        \hline
         Model & Sequence Matcher &  Levenshtein Distance & Jaccard Index & Cosine Similarity \\
         \hline
         GPT 4o Mini & 23.31 & 43.62 & 86.79 & 81.84 \\
         \hline
         GPT 3.5 & 33.02 & 46.68 & 84.22 & 81.51 \\
         \hline
         Gemini & 12.31 & 34.24 & 83.9 & 79.69 \\
         \hline
         Cohere & 10.6 & 31.9 & 73.87 & 72.57 \\
         \hline
         Llama3 & 13.86 & 32.09 & 80.63 & 73.39 \\
         \hline
    \end{tabular}

\end{table}

\begin{table}[!h]
\caption{Difference between the average Similarity Scores for Information vs Situation Questions}
    \label{tab:sim_tab_diff}
    \begin{tabular}{|p{1.5cm} | p{1.5cm} | p{1.5cm} | p{1.3cm} | p{1.1cm} |}
\hline
Model           & Sequence Matcher & Levenshtein Distance & Jaccard Index & Cosine Similarity \\ \hline
GPT 4o Mini & 6.9              & 2.92                 & 2.96          & 7.45              \\ \hline
GPT 3.5     & -2.2             & 3.88                 & 5.27          & 3.12              \\ \hline
Gemini   & -2.09            & -1.74                & 2.72          & 2.41              \\ \hline
Cohere          & 2.73             & 1.45                 & 5.58          & 8.46              \\ \hline
Llama3          & 0.34             & 1.79                 & 4.34          & 7.85              \\ \hline
\end{tabular}
\end{table}

\subsubsection{How Types of Questions affect LLM Consistency}
When looking at the raw similarity scores for each question for each LLM, we noticed some patterns. The question "What are the response codes that can be received from a Web Application?" yielded the highest similarity scores for 3 out of 4 metrics on GPT 4o Mini, 1 out of 4 metrics on GPT 3.5, and 3 out of 4 metrics on Gemini. Interestingly, this same question yielded the lowest similarity score on 2 out of 4 of the metrics for Cohere, specifically Sequence Matcher and Levenshtein Distance, the two metrics that take order into account. On the other two metrics, that question scored fairly high for Cohere. It scored a little less than Cohere's highest recorded Jaccard index and in the middle of its highest and lowest recorded Cosine similarity scores. Looking at the responses themselves, they provide the same explanations, but with varying degrees of elaboration. This shows Cohere is not factually inconsistent but still inconsistent in its responses. Overall, there has been very little variation in Cosine and Jaccard scores, which shows relative factual consistency, but a lot of variation in Sequence Matcher and Levenshtein scores which shows that all the LLMs studied display this to some degree.

If we use Sequence Matcher and Levenshtein Distance scores to represent consistency in explanations and Jaccard Index and Cosine Similarity scores to represent informational consistency, looking at a table of the differences between the average Similarity Scores
for Information vs Situation Questions (Table \ref{tab:sim_tab_diff}), informational consistency drops much more for the situation questions than consistency in explanation. In some cases, the average Sequence Matcher and Levenshtein Distance scores are even higher for situational questions, most notably with Gemini, where they are both higher. This shows that LLMs are less factually accurate when given more open to interpretation situational questions, but are sometimes more consistent in explaining their responses to those questions.

\subsection{Agreement}
For agreement, Meta OPT~\cite{metaopt}'s responses had to be left out, because it refuses to give a yes or no answer on whether a response is correct or not, usually responding with "I think it's a good question", or just answering the question again, no matter what prompt we used. This means it cannot be used for Self-Validation, and cannot Cross-Validate the other models. So it isn't included in the Self-Validation plots, and on the Cross-Validation plots every other model only has 5 other LLMs to check its answer instead of 6, Meta OPT is not included.

\subsubsection{Self-Validation}
\begin{figure}[!tb]
    \centering
    \includegraphics[width=\linewidth]{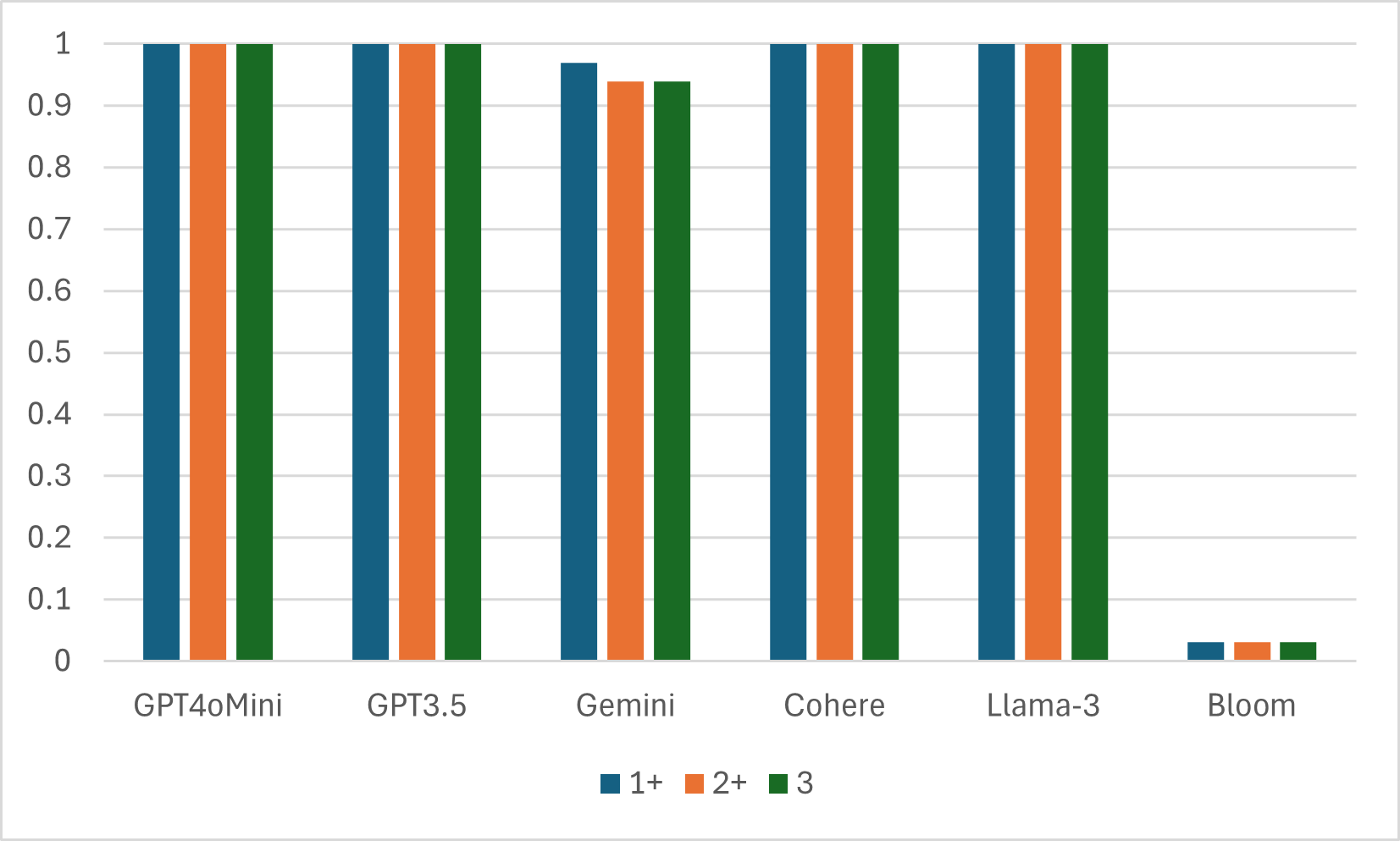}
    \caption{Self Validation for Information Questions}
    \label{fig:self-valid}
\end{figure}

\begin{figure}[!tb]
    \centering
    \includegraphics[width=\linewidth]{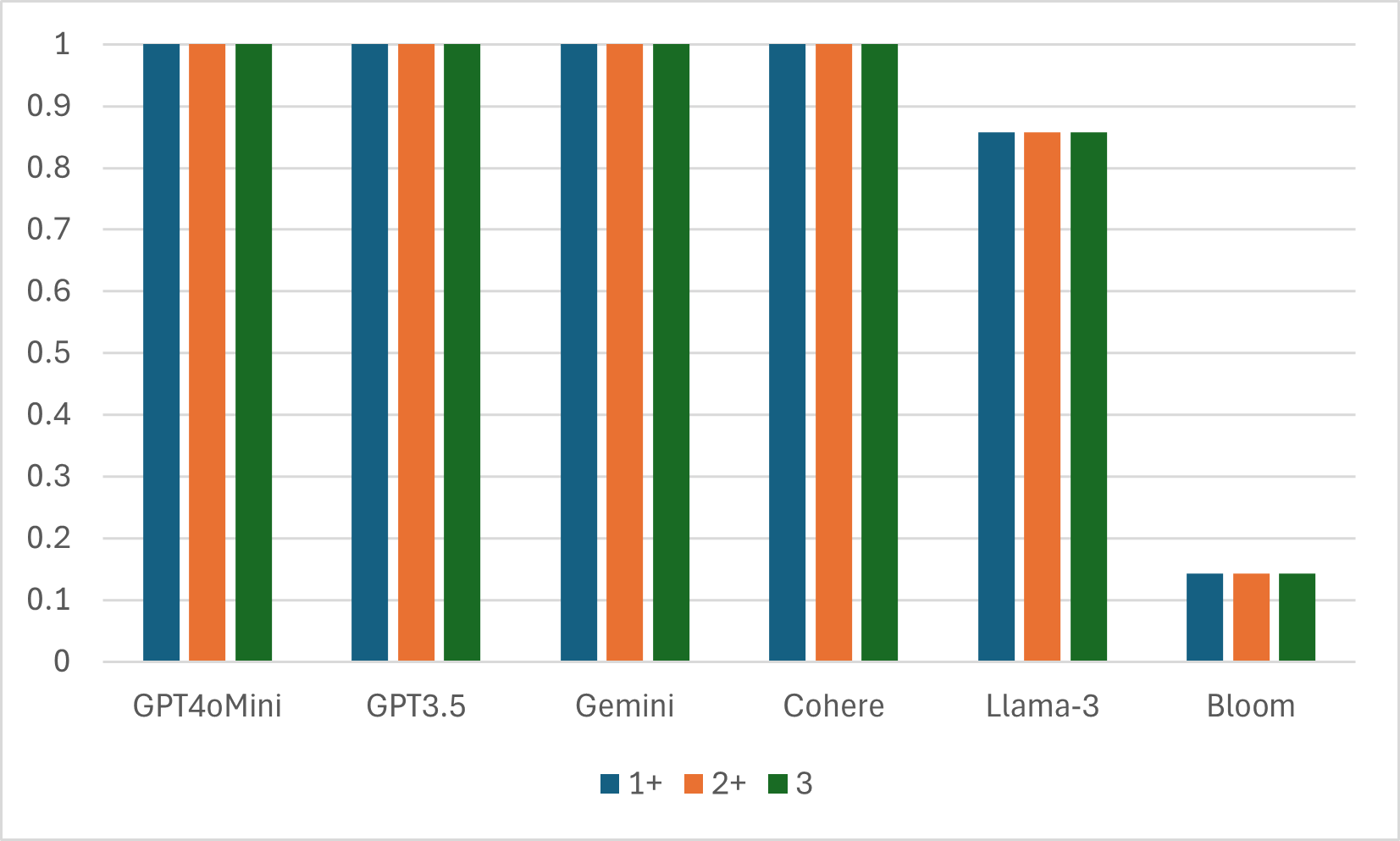}
    \caption{Self Validation for Situational Questions}
    \label{fig:self-valid-sit}
\end{figure}

For Self-Validation, whether it's informational (Figure \ref{fig:self-valid}) or situational questions (Figure \ref{fig:self-valid-sit}), most of the models agree with themselves more than 80\% of the time. Bloom is the only model that is critical of itself. Interestingly, it appears to agree with itself more on the situational questions than the informational questions.

\subsubsection{Cross-Validation}
For Cross-Validation, the plots are organized by the number of LLMs that agree with one LLM's response. The first bar depicts the case when 2 of the 5 LLMs agree with a response, and the last bar depicts the case when all 6 agree. The last bar is only applicable for Meta OPT since Meta OPT itself cannot be used to cross-validate the other models, they all have 5 models checking their responses, while Meta OPT has all 6.

\begin{figure}[!tb]
    \centering
    \includegraphics[width=\linewidth]{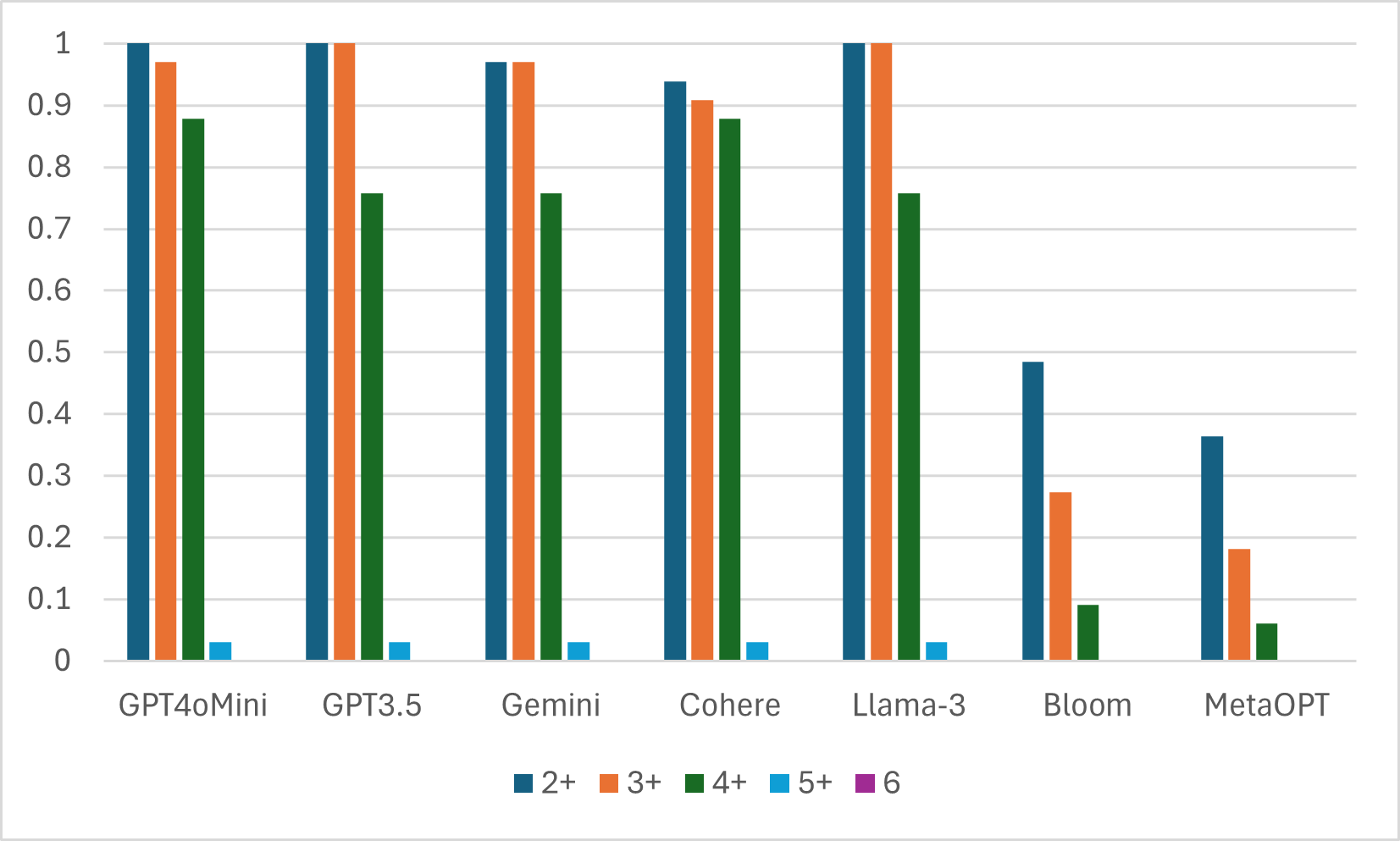}
    \caption{Cross Validation for Information Questions}
    \label{fig:cross-valid}
\end{figure}

\begin{figure}[!tb]
    \centering
    \includegraphics[width=\linewidth]{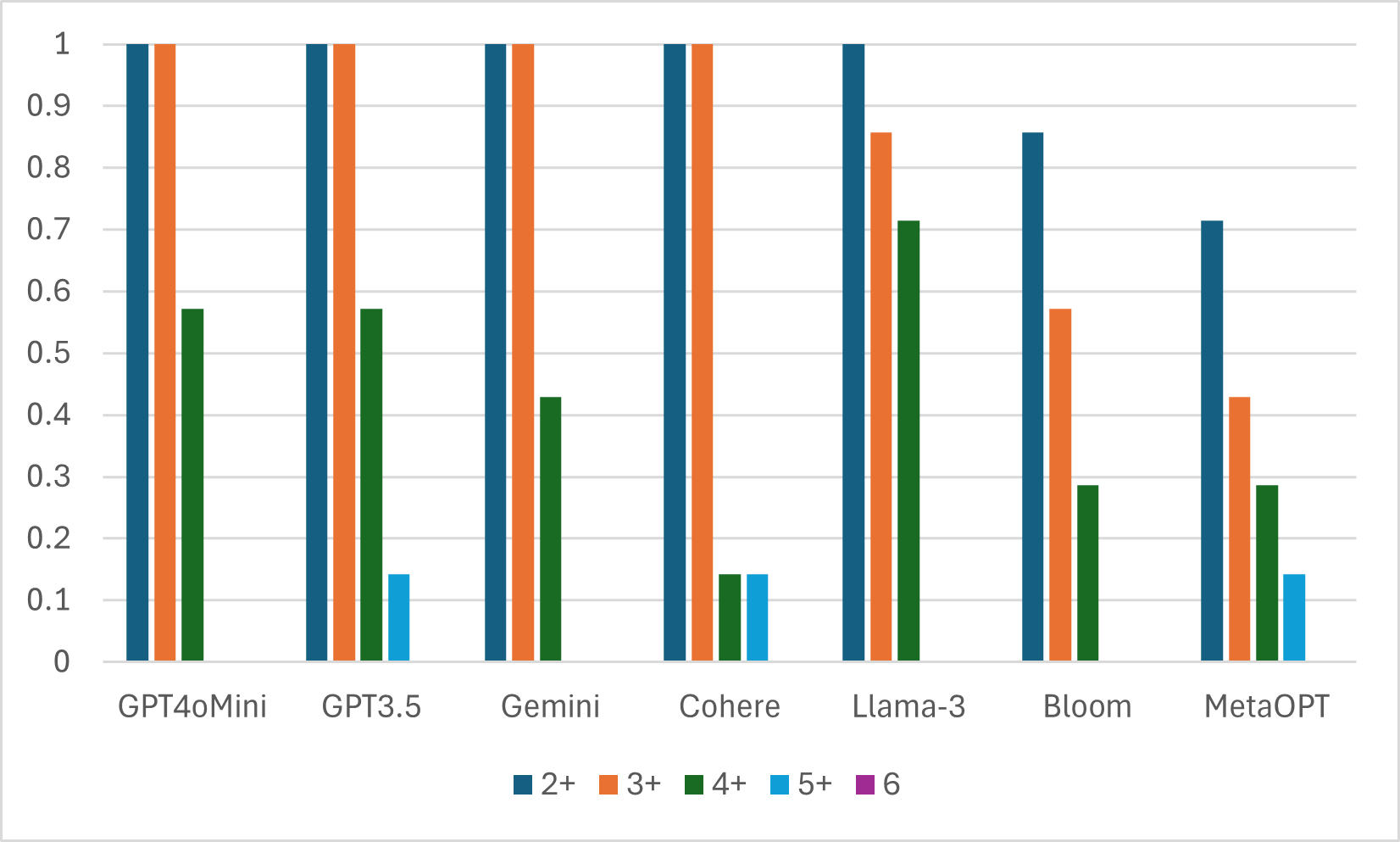}
    \caption{Cross Validation for Situational Questions}
    \label{fig:cross-valid-sit}
\end{figure}

Interestingly, despite being the only 100\% consistent models, Meta OPT and Bloom score the lowest in both the informational questions (Figure \ref{fig:cross-valid}) and the situational questions (Figure \ref{fig:cross-valid-sit}). They actually score higher on the situational questions than the informational ones. Along with Gemini, they're the only models to behave this way.

\subsubsection{Model Hallucination and Validation} \label{sec:hallucination}
When asked to "Explain CIA triad in Cybersecurity", Meta OPT provided an incorrect answer, talking about the Central Intelligence Agency instead of the CIA triad. In the Cross-Validation step, 4 of the 6 LLMs responded that the answer was incorrect, which marks the overall question as incorrect by our parameters. While that showcases the strengths of Cross-Validation with other LLMs, two LLMs still got the validation wrong, those being GPT 3.5 and Cohere. More interestingly, Cohere answered "Yes, you explained the CIA triad, which stands for confidentiality, integrity, and availability, in cybersecurity", generating the correct answer itself but wrongfully believing it was provided in the prompt. Future work on refining the effectiveness of the Cross-Validation algorithm could be on marking the LLMs which are most accurate when validating answers and removing those that are the least. 

Detecting a hallucination is more difficult with situational questions. When prompted with "A friend of yours sends an e-card to your mail. You have to click on the attachment to get the card. What do you do? Justify your answer", Meta OPT simply responded with "I have to click on the attachment to get the card", where the correct answer would be to confirm that it is not malicious first. In the cross-validation step, only GPT 4o Mini identified this as the incorrect answer. Despite Gemini, GPT 3.5, Cohere and Llama3 all giving the correct answer when directly responding to the question, they still believed the incorrect answer was correct despite correctly identifying the security risks in their own responses. This shows the weakness in using LLM agreement to double-check responses for more abstract questions, as an LLM's response to a situation can be a possible course of action, but not necessarily a correct one.


\section{Related Work} \label{sec:related}
Our paper focuses on how to analyze the consistency of responses by an LLM. Hamman et al~\cite{hamman2024quantifying} defines the measure of consistency for models predicting tabular data. Our work relies on general large language models that are not specifically fine-tuned for tabular data.  

Consistency analysis is an important aspect of hallucination analysis and detection. There is an entire body of work on hallucination~\cite{liu2024exploring}. There are other works related to self-reflection~\cite{ji-etal-2023-towards}, distributed LLMs~\cite{10476280}, and mixture-of-experts~\cite{du2024mixture}. \cite{jang2023consistency} performed experiments to examine the consistency of ChatGPT's~\cite{openai2023gpt35, openai2023gpt4} responses, uncovering situations where its language comprehension abilities don't always lead to logically sound predictions. 

 LLMs suffer from security issues~\cite{wu2024new,qachfar2024all}, and there are risks of code generation using LLMs~\cite{vaidya2023critical}. LLMs hallucinate~\cite{xu2024hallucination} and such hallucinations often increase the risk of reduction of utility of the models, or even risk of failure during task execution, with LLMs as judges~\cite{zheng2024judging}, or autonomous LLM agents~\cite{talebirad2023multi}.

LLMs can be used in enabling cybersecurity processes (SecOps)~\cite{gennari2024considerations}. Aljanabi et al.~\cite{aljanabi2023chatgpt} explored ChatGPT's relevance in the cybersecurity domain, demonstrating its utility for security professionals in identifying potential vulnerabilities. Khoury et al.~\cite{khoury2023secure} conducted experiments to assess the safety of programs generated by ChatGPT and sought security assessments. Peng et al.~\cite{peng2023check} introduced a new model that enhances black-box LLMs with plug-and-play modules, outlining an iterative strategy to improve model feedback. LLMs are also being proposed to be used in security operation centers (SoC)~\cite{saha2024llm}. \par

Secure code copilots and secure generation of code is an important area of investigation, in security analysis of code and configurations~\cite{toth2024llms, minna2024analyzing}, in generation of secure code~\cite{vaidya2023critical,saha2024empowering}, in security and penetration testing~\cite{song2024poster}. 

Existing studies have focused on the broader implications and uses of ChatGPT~\cite{lund2023chatting}, exploring the transformative potential of ChatGPT in academia and libraries, and discussing its benefits for search, discovery, cataloging, and information services. Meanwhile, discussions on the impact on higher education~\cite{rudolph2023chatgpt} highlight concerns about its ability to generate text that resembles human-authored content.\cite{baidoo2023education} examined the pros and cons of incorporating ChatGPT into teaching and learning, adding to discussions about educational technology, while~\cite{jiao2023chatgpt} investigates how well ChatGPT can translate across multiple languages. ~\cite{kuchnik2023validating} presents an innovative technique for validating and querying LLMs using standard regular expressions. Liu et al.~\cite{liu2023summary} conducted a thorough study that included trend analysis, word cloud representations, and domain-wise distribution analysis. This investigation shed light on the model's capabilities, ethical considerations, and future directions. \par

\section{Conclusions and Future Work} \label{sec:conclusions}
In this paper, we ask the following question -- "how consistent are LLM responses, both in the information provided and factually" especially in the context of their use in cybersecurity. LLMs are expected to be used for various security operations in the industry~\cite{gennari2024considerations}. Before we put significant reliance on the black-box LLMs (which the industry has already started), can we evaluate such models on how consistent they are in their responses, and can security stakeholders such as CISOs make decisions on whether and how to use which LLM for tasks as important as cybersecurity operations?

We have carried out an extensive set of experiments and analyzed the consistency of LLMs for their responses against a benchmark of cybersecurity questions. Our experiments demonstrate that LLMs have made significant strides in improving consistency and reducing hallucination in the past couple of years, as newer models like GPT 4o Mini and Meta Llama3 outperform older ones like Meta OPT and Bloom. Despite that, LLMs still have quite a way to go before they become usable for important cybersecurity operations. When confronted with more abstract situational questions, there is a clear drop in consistency and agreement between LLMs. While our self and cross-validation algorithms have been effective at detecting LLM hallucination, they become less reliable the more abstract a question is. In the future, we plan to conduct a more supervised analysis of how accurate LLMs are to specifically select the ones with the best track record for response validation.

We plan to explore the relations between consistency and hallucination in detail, as well as carry out further experiments on classifying LLMs to different cybersecurity tasks. Further understanding of how inconsistent are the responses and how they are generated based on an analysis of the internal states of the models and attention layers may help us fine-tune the models better.

\bibliographystyle{IEEEtranS}
\bibliography{references}
\vspace{12pt}

\newpage

\section*{Appendix}

\begin{table}[!h]
\caption{List of Cybersecurity Information Questions}\label{tbl:infq}
\begin{tabular}{lp{3.2in}}
Q1:  & What is the difference between   VA(Vulnerability Assessment) and PT (Penetration Testing)? \\
Q2:  & What is a three-way handshake?                                                             \\
Q3:  & What are the response codes that can be received from a Web Application?                   \\
Q4:  & What is the difference between HIDS and NIDS?                                              \\
Q5:  & What are the steps to set up a firewall?                                                   \\
Q6:  & Explain SSL Encryption                                                                     \\
Q7:  & What steps will you take to secure a server?                                               \\
Q8:  & Explain Data Leakage                                                                       \\
Q9:  & What is Port Scanning?                                                                     \\
Q10: & What are the different layers of the OSI model?                                            \\
Q11: & What is a VPN?                                                                             \\
Q12: & What do you understand by Risk, Vulnerability \& Threat in a network?                      \\
Q13: & How can identity theft be prevented?                                                       \\
Q14: & What are black hat, white hat, and grey hat hackers?                                        \\ Q15: & How often should you perform Patch management?                                             \\
Q16: & How would you reset a password-protected BIOS configuration?                               \\
Q17: & What is an ARP and how does it work?                                                       \\
Q18: & What is port blocking within LAN?                                                          \\
Q19: & What are salted hashes?                                                                    \\
Q20: & Explain SSL and TLS                                                                        \\
Q21: & What is 2FA and how can it be implemented for public websites?                             \\
Q22: & What is Cognitive Cybersecurity?                                                           \\
Q23: & What is the difference between VPN and VLAN?                                               \\
Q24: & What is cryptography?                                                                      \\
Q25: & What is the difference between VPN and VLAN?                                               \\
Q26: & What is the difference between Symmetric and Asymmetric encryption?                        \\
Q27: & What is the difference between IDS and IPS?                                                \\
Q28: & Explain the CIA triad in cybersecurity.                                                        \\
Q29: & How is Encryption different from Hashing?                                                  \\
Q30: & What is a Firewall and why is it used?                                                     \\
Q31: & What are some of the common Cyberattacks?                                                  \\
Q32: & What protocols fall under the TCP/IP internet layer?                                           \\
Q33: & What is a Botnet?                                                                         
\end{tabular}

\vspace{0.3cm}
\caption{List of Cybersecurity Situation Questions}\label{tbl:sitq}

\begin{tabular}{lp{3.2in}}
Q1: & A friend of yours sends an e-card to your mail. You have to click on the attachment to get the card. What do you do?   Justify your answer                                                                                                                                                                                                                                                  \\
Q2: & In our computing labs, print billing is often tied to the user’s login.   Sometimes people call to complain about bills for printing they never did only to find out that the bills are, indeed, correct. What do you infer from this situation? Justify                                                                                                                                  \\
Q3: & Which of the following passwords meets UCSC’s password requirements?   a).@\#\$)*\&\textasciicircum{}\% b).akHGksmLN c).UcSc4Evr! d).Password1                                                                                                                                                                                                                                                \\
Q4: & You receive an email from your bank telling you there is a problem with your account. The email provides instructions and a link so you can log into your account and fix the problem. What should you do?                                                                                                                                                                                \\
Q5: & A while back, the IT folks got a number of complaints that one of our campus computers was sending out Viagra spam. They checked it out, and the reports were true: a hacker had installed a program on the computer that made it automatically send out tons of spam emails without the computer owner’s knowledge. How do you think the hacker got into the computer to set this up? \\
Q6: & There is this case that happened in my computer lab. A friend of mine used their Yahoo account at a computer lab on campus. She ensured that her account was not left open before she left the lab. Someone came after her and used the same browser to re-access her account. and they started sending emails from it. What do you think might be going on here?                     \\
Q7: & Two different offices on campus are working to straighten out an error in an employee’s bank account due to a direct deposit mistake. Office \#1 emails the correct account and deposit information to Office \#2, which promptly fixes the problem. The employee confirms with the bank that everything has,   indeed, been straightened out. What is wrong here?                     
\end{tabular}

\end{table}

\end{document}